\documentclass[article,balancelastpage,twocolumn,prl]{revtex4}%
\usepackage{makeidx}
\usepackage{amssymb}
\usepackage{dcolumn}
\usepackage{graphicx}
\usepackage{acronym}
\usepackage{amsmath}
\usepackage{amsfonts}%
\ifx\pdfoutput\relax\let\pdfoutput=\undefined\fi
\newcount\msipdfoutput
\ifx\pdfoutput\undefined\else
\ifcase\pdfoutput\else
\ifx\paperwidth\undefined\else
\ifdim\paperheight=0pt\relax\else\pdfpageheight\paperheight\fi
\ifdim\paperwidth=0pt\relax\else\pdfpagewidth\paperwidth\fi

\begin{document}	
	
\title{Cylindrical symmetric solutions of the Einstein equation with the cosmological term}
\author{B. V. Gisin }
\affiliation{E-mail: borisg2011@bezeqint.net}
\date{\today }

\begin{abstract}
Assuming that universe is the object of point rotation at a frequency,
the relationship is established between this frequency and the cosmological
constant. Using the transformation for point-like rotating coordinate systems, an unusual exact solution of the Einstein equation is found.


\end{abstract}
\maketitle

\section{Introduction}

Recently, a three-dimensional transformation was obtained by successive rotations of three pairs of differentials of cylindrical coordinates, an angle, a distance along the rotation axis and time \cite{pr}. 

The transformation, obtained within the framework of a standard physical approach, has two important properties: the turning out into the Lorentz transformation if the rotation frequency tends to zero, and the invariance of the quadratic form composed from these three differentials. 

Simultaneously, interval of the flat Minkowski space is invariant under the transformation, since it contains this quadratic form. 

This is means that the space can be the object of a point rotation \cite{pr}. An example of such a rotation is the circularly polarized electromagnetic wave where the axis of rotation exists at each point.

It is well known that the search for solutions to the Einstein equation is the subject of some conjectures about the form of the metric. A typical example is the Schwarzschild solution, in which the following assumptions are used: the metric is diagonal, the radial and time components depend only on the spherical radius, and the remaining two components coincide with their Euclidean values.

In the paper, we are looking for cylindrical solutions, assuming that the interval in the general theory of relativity also contains the above quadratic form.

We consider free space without invoking the energy-momentum tensor. Partially because of the Einstein note "Our problem now is introduce a tensor $T_{\mu\nu}$, of the second rank, whose structure we do know but provisionally..." \cite{E}. The problem is behind of the scope of the paper. 
 
Unlike the Schwarzschild metric, static solutions that depend only on the cylindrical radius do not exist, whereas an unusual solution is found that depends on the cylindrical angle, distance along the axis of rotation and time.

\section{The Einstein equation}

The most common form of the interval is 
\begin{align}
 ds^{2}=g_{\mu\nu}dx^{\mu}dx^{\nu}=gdr^{2}+f(r^{2}d\varphi^{2}+dz^{2}-dt^{2}),\label{ds}%
\end{align}
where $g_{\mu\nu}$ is the metric tensor, $g,f$ are some functions of coordinates. 

The quadratic form in brackets is an invariant with respect to the 3D transformation.

The co-and contravariant components of the metric tensor (\ref {ds}) are easily determined
\begin{align}
& g_{rr}=g, \;\; g_{\varphi\varphi}=r^2 f, \;\; g_{zz}=f, \;\; g_{tt}=-f, \\
& g^{rr}=\frac{1}{g}, \;\; g^{\varphi\varphi}=\frac{1}{r^2 f}, \;\; g^{zz}=\frac{1}{f}, \;\; g^{tt}=-\frac{1}{f}. \label{cgmn}
\end{align}

The Einstein equation with the cosmological term in the free space can be written as
\begin{align}
R_{\mu\nu}=g_{\mu\nu}\Lambda,\label{E0}%
\end{align} 
where $R_{\mu\nu}$ is the Ricci tensor
\begin{align} 
& R_{\mu\nu}=\Gamma^{\sigma}_{\mu\nu,\sigma}-\Gamma^{\sigma}_{\mu\sigma,\nu}+
\Gamma^{\sigma}_{\mu\nu}\Gamma^{\rho}_{\sigma\rho}- \Gamma^{\sigma}_{\mu\rho}\Gamma^{\rho}_{\nu\sigma}, \label{Ricc} \\ 
& \Gamma^{\sigma}_{\mu\nu} = \frac{1}{2}g^{\sigma\rho}(g_{\mu\rho,\nu}+g_{\rho\nu,\mu}-g_{\mu\nu,\rho}),\label{Gam}
\end{align}
$\Lambda$ is the cosmological constant, $\Gamma^{\sigma}_{\mu\nu}$ is the Christoffel symbol.

The Ricci tensor contains ten components, six of which are nondiagonal and four-diagonal. 

The metric tensor contains two independent functions $ f, g $. Surprisingly, all ten Einstein equations (\ref{E0}) are compatible for these two functions $ f, g $.

\section{Solutions}

\subsection{Nondiagonal components}
	
It can be straightforwardly shown that Eq. (\ref{E0}) has no static solutions with functions $f,g$ depending only on $r$. Perhaps this statement can be proved for the general case, ie, the dependence of the metric tensor on all the coordinates.

In any case, we assume here that the metric tensor does not depend on $r$.

We start from nondiagonal components of the Ricci tensor. 

If $f_{,r}=0, \; g_{,r}=0$, then the equation $R_{r\varphi}=0$ turns out into
\begin{align}
f_{,\varphi}=0. \label{Rv}
\end{align}

From  $R_{rz}=0, \; R_{rt}=0$ one follows
\begin{align}
\frac{g_{,z}}{g}=\frac{f_{,z}}{f}, \quad \frac{g_{,t}}{g}=\frac{f_{,t}}{f}. \label{Rrzrt}
\end{align}
The two equations are equivalent to the relation
\begin{align}
g=Cf, \label{gCf}
\end{align}
where $f$ depends only on $z,t$ while $C$ depends only on $\varphi$.

Because of relation (\ref{gCf}), the equations $R_{\varphi z}=0$ and $R_{\varphi t}= 0$ are automatically satisfied.

The equation $R_{zt}=0$ can be transformed to the form $(f^{-1/2})_{,zt}=0$.
Solution of this equation corresponds to the sum of two functions  $f_z$ and $f_t$ depending on $z$ and $t$ respectively. Finally the form of $f$ is
\begin{align}
f=\frac{1}{(f_z+f_t)^2}. \label{fzt}
\end{align}

\subsection{Diagonal components}

Now we arrive at equations for the diagonal components, excluding terms that depend on $ r $ and $ g $ with the help of (\ref {gCf}). The equation $R_{rr}=g_{rr}\Lambda$ can be written then as
\begin{align} 
& \frac{f}{g}R_{rr}:-\frac{C_{,\varphi\varphi}}{2r^2C}+\frac{C^2_{,\varphi}}{4r^2C^2}
-\frac{f_{,zz}}{2f}+\frac{f_{,tt}}{2f}=f\Lambda. \nonumber
\end{align}
Terms with $C$ depend on $r$ and $\varphi$ while terms with $f$ depend on $z,t$. Separating variables, this equation can be written as two independent equations
\begin{align} 
& -\frac{C_{,\varphi\varphi}}{C}+\frac{C^2_{,\varphi}}{2C^2}=0, \quad C=(\eta\varphi)^2, \label{Ceq} \\ %
& -\frac{f_{,zz}}{2f}+\frac{f_{,tt}}{2f}=f\Lambda. \label{RCrr}
\end{align}
Solution of the first equation can be easily found. In this solution we discard the integration constant that determines the initial angle. 

For three  remaining equation we obtain after above changes
\begin{align} 
& \frac{1}{r^2}R_{\varphi\varphi}: -\frac{f_{,zz}}{2f}+\frac{f_{,tt}}{2f}=f\Lambda,
\label{Rvv} \\
& R_{zz}: -\frac{3f_{,zz}}{2f}+\frac{3f^2_{,z}}{2f^2}+\frac{f_{,tt}}{2f} =f\Lambda. \label{Rzz} \\ 
& R_{tt}: -\frac{3f_{,tt}}{2f}+\frac{3f^2_{,t}}{2f^2}+\frac{f_{,zz}}{2f}=-f\Lambda. \label{Rtt} 
\end{align}
The equation (\ref{Rvv}) coincides with (\ref{RCrr}).

Subtract  equations (\ref{Rvv}) and (\ref{Rzz})
\begin{align} 
\frac{f_{,zz}}{f}-\frac{3f^2_{,z}}{2f^2}=0, \;\; (f^{-1/2})_{,zz}=0, 
\nonumber
\end{align}
Solution of this equation in comparison with (\ref{fzt}) allows us to conclude that $f$ and $g$ has the form
\begin{align} 
f=\frac{1}{(\nu t+kz)^2}, \quad g=\frac{(\eta\varphi)^2}{(\nu t+kz)^2},\label{fnu}
\end{align}
where $\nu,k$ are constants. Without loss generality we drop the integration constant.

This solution satisfies all the equations provided that 
\begin{align} 
\nu^2-k^2=\frac{1}{3}\Lambda. \label{nk}
\end{align}

$f$ and $g$ have the singularity at $t=0,\; z=0$. In the course of time, this singularity moves along the $z$-axis
\begin{align} 
z=-\frac{\nu}{k}t. \label{zt}
\end{align}
However, the possibility of periodic motion with a return to its original position should not be excluded in the more general case with the energy-momentum tensor.

\section{Discussion}

Two type of rotation exist in nature. The first is a mechanical type with one rotation axis. An example of the second type is a circularly polarized electromagnetic wave, where the rotation axis exists at each point. For such a rotation we use the term 'point rotation'.

Obviously, the solution presented above belongs to the second type.

\subsection{Rotating universe}

In \cite{pr} it was assumed that our universe could be the object of the point rotation, because the flat Minkowski spacetime is invariant with respect to the three-dimensional transformation.

Assume that the rotation frequency of universe is $\omega_r$, named in \cite{pr} relic frequency, and the corresponding wavelength is $\lambdabar_{r} \equiv c/\omega_r$.  These are characteristic constants in the same sense as the speed of light.
The 'cosmological' normalization of coordinates in both the frames
\begin{align}
\frac{r}{\lambdabar_r} \rightarrow r, \;\; \frac{z}{\lambdabar_r} \rightarrow z, \;\; \frac{ct}{\lambdabar_r} \rightarrow t, \label{N}
\end{align}
removes $\omega_r$ and $\lambdabar_r$ from the transformation, as well as from any relativistic invariant equation including the Einstein equation (\ref{E0}).
The scalar curvature corresponding this equation is $R=4\Lambda$ 

The Ricci tensor contains the second derivative and the product of two first derivatives of the metric tensor with respect to the coordinates. Therefore, the normalization of (\ref {E0}) is realized by multiplication by 
$\lambdabar^2_r $.

\subsection{Cosmological constant}

A particular (but not rigorous) example of the relationship between $\lambdabar_r $ and $ \Lambda $ can be easily found if the space is a hypersurface in a 5D-cylindrical space of radius $\lambdabar_r $.
In this case $R=2/\lambdabar^2$ and this relationship is
\begin{align} 
\Lambda=\frac{1}{2\lambdabar^2}. \label{Ll}
\end{align}  
From this point of view, the cosmological term is nothing but a consequence of  the universe point rotation.

The speed of light constancy by the motion around a circle of radius $r$ determines the allowable value of $r$. For frequency $\omega_r$ this means
\begin{align} 
\frac{r\omega_r}{c}\leq 1, \;\; \text{or} \;\; r\leq\lambdabar_r \label{rom}
\end{align}   
 
If the association (\ref{Ll}) is true then $2\lambdabar_{r}$ determines the size of universe.

It is obvious that at the present time the huge period of rotation of the
universe about $6.64 \cdot10 ^{15} $ years \cite{Tan} does not affect ordinary life.
 
\section{Conclusion}

In the cylindrical solution given here, a rotating universe in free space is described.
In such an universe the cosmological constant is proportional to the square of the rotational frequency. 

For some values $t$ and $z$ the metric tensor tends to infinity. This 'state of  the Big Bang for the free space' moves along the axis of rotation with the passage of time.

Beyond these values, for any time interval that is much shorter than the period of  rotation and the length of interval on the z axis, which is much smaller than $\lambdabar_{r} $, the space can be interpreted as a flat Minkowski space.

This approach contains a dualism, analogous to the concept of 'wave-particle' in quantum theory. On the one hand, the universe has a size of $2\lambdabar_{r}$, on the other hand, the universe is infinite, because the rotation axis, as the reference axis, can be chosen at each point.


\begin{thebibliography}{9}                                                                                                
\bibitem {pr}B. V. Gisin, \emph{Is our space an object of point rotation?},
arXiv: 1712.06964v1 [physics.gen-ph] 15 Dec 2017.

\bibitem {E} A. Einstein, \emph{The Meaning of Relativity}, University Press
(1970), Fifth edition, Princeton, N. J.

\bibitem {Tan}M. Carmeli and T. Kuzmenko, \emph{Value of the Cosmological
Constant: Theory versus Experiment}, 2001, https://cds.cern.ch/record/485959/files/0102033.pdf.

\end{thebibliography}
\end{document}